\documentclass[preprint,showpacs,preprintnumbers,amsmath,amssymb]{revtex4}
\usepackage{booktabs}
\usepackage{mathrsfs}
\usepackage{epsfig}
\usepackage{graphicx}
\usepackage{dcolumn}
\usepackage{bm}
\usepackage{amsmath}
\usepackage{slashed}       
\usepackage{amssymb}
\usepackage{amsfonts}

\let\jnfont=\rm
\def\NPB#1,{{\jnfont Nucl.\ Phys.\ B }{\bf #1},}
\def\PLB#1,{{\jnfont Phys.\ Lett.\ B }{\bf #1},}
\def\EPJC#1,{{\jnfont Eur.\ Phys.\ Jour.\ C }{\bf #1},}
\def\PRD#1,{{\jnfont Phys.\ Rev.\ D }{\bf #1},}
\def\PRL#1,{{\jnfont Phys.\ Rev.\ Lett.\ }{\bf #1},}
\def\MPLA#1,{{\jnfont Mod.\ Phys.\ Lett.\ A }{\bf #1},}
\def\JPG#1,{{\jnfont J.\ Phys.\ G}{\bf #1},}
\def\CTP#1,{{\jnfont Commun.\ Theor.\ Phys.\ }{\bf #1},}
\def\ZPC#1,{{\jnfont Z.\ Phys.\ C }{\bf #1},}
\def\JHEP#1,{{\jnfont JHEP \ }{\bf #1},}
\def\lsim{\raise0.3ex\hbox{$<$\kern-0.75em\raise-1.1ex\hbox{$\sim$}}}
\def\gsim{\raise0.3ex\hbox{$>$\kern-0.75em\raise-1.1ex\hbox{$\sim$}}}

\newcommand{\8}{\ \ \ \ \ \ \ \ }

\begin{document}
\preprint{\parbox{1.2in}{\noindent arXiv:1312.4260}}

\title{Pair production of 125 GeV Higgs boson in the SM extension
with color-octet scalars at the LHC}

\author{Zhaoxia Heng$^1$, Liangliang Shang$^1$, Yanming Zhang$^1$, Yang Zhang$^{1,2}$, Jingya Zhu$^2$}

\affiliation{
  $^1$   Department of Physics,
        Henan Normal University, Xinxiang 453007, China \\
  $^2$   State Key Laboratory of Theoretical Physics,
      Institute of Theoretical Physics, Academia Sinica, Beijing 100190, China
      \vspace{1cm}}

\begin{abstract}
Although the Higgs boson mass and single production rate have been determined
more or less precisely, its other properties may deviate significantly
from its predictions in the standard model (SM) due to the uncertainty of
Higgs data. In this work we study the Higgs pair production at the LHC
in the Manohar-Wise model, which extends the SM by one family of
color-octet and isospin-doublet scalars.
We first scanned over the parameter space of the Manohar-Wise model
considering exprimental constraints
and performed fits in the model to the latest Higgs data by using
the ATLAS and CMS data separately.
Then we calculated the Higgs pair production rate
and investigated the potential of its discovery at the LHC14.
We conclude that:
(i) Under current constrains including Higgs data after Run I of the LHC,
  the cross section of Higgs pair production in the Manohar-Wise model
  can be enhanced up to even $10^3$ times prediction in the SM.
(ii) Moreover, the sizable enhancement comes from the contributions
  of the CP-odd color-octet scalar $S^A_I$. For lighter scalar $S^A_I$ and
  larger values of $|\lambda_I|$, the cross section of Higgs pair production can be much larger.
(iii) After running again of LHC at 14 TeV, most of the parameter spaces in the Manohar-Wise model can be test.
  For an integrated luminosity of 100 fb$^{-1}$ at the LHC14,
  when the normalized ratio $R=10$, the process of Higgs pair production can be detected.
\end{abstract}

\pacs{14.80.Bn, 12.60.Fr}

\maketitle

\section{Introduction}
In July 2012, both the ATLAS and CMS collaborations at the LHC
announced the discovery of a new boson with mass around
125 GeV \cite{1207atlas,1207cms}. The combined data
at the LHC indicate that its properties are quite compatible
with those of the Higgs boson in the Standard Model (SM) \cite{13atlas,13cms}.
However, whether the new boson is the Higgs boson predicted by the SM or
new physics models still need to be further confirmed by the LHC experiment
with high luminosity.
So far, various new physics models like the low energy supersymmetric models
can give reasonable interpretations for the properties of this SM-like Higgs boson around 125 GeV \cite{125-our,125-other-MSSM,natural-susy-125,NMSSM-125,cmssm-125}.

Moreover, discovery of the SM-like Higgs boson is not the end of the story.
The next challenge for the experiment is to precisely measure its properties
including all the possible production and decay channels.
As a rare production channel, the Higgs pair production can be
used to test the Higgs self-couplings effectively \cite{self-coupling},
which play an essential role in reconstructing the Higgs potential.
The Higgs pair production in the SM at the LHC proceeds through the gluon fusion $gg\to hh$.
At the leading order, the main contributions come from the heavy quark loops
through the box diagrams and triangle diagrams with the Higgs self-coupling.
Due to the weak Yukawa couplings and Higgs self-coupling, as well as
the cancelations between these two types of diagrams,
the cross section in the SM is too small to be detected with current integrated luminosity.
Even at $\sqrt{s} =$ 14 TeV with high luminosity, it is still difficult to
detect this process. The discovery potential of the LHC to detect this production process
has been investigated in \cite{hh-detect,Barger:2013jfa,hh-other},
and the most promising channel to detect it is $gg\to hh\to b \bar b \gamma\gamma$,
other signal channels such as $h h \to b \bar{b} \tau^+ \tau^-$ are swamped by
the reducible backgrounds \cite{Barger:2013jfa}.

Compared with the predictions in the SM, the production rate of the SM-like Higgs pair
production in new physics models can be enhanced significantly
due to relatively large additional couplings of the SM-like Higgs boson
with the introduced new particles, such as squarks in supersymmetric models \cite{hh-susy}
or other colored particles \cite{color-scalar}.
Therefore, the Higgs pair production can be a sensitive probe to new physics beyond the SM.
In this paper we investigate the effects of color-octet scalars in the
Manohar-Wise (MW) model \cite{MW-Model} on the Higgs pair production at the LHC.
The Manohar-Wise model is a special type of two-Higgs-doublet model and predicts a
family of color-octet scalars, which can have sizable couplings with the Higgs boson,
since the sign of Higgs coupling with gluons is usually opposite to the prediction
in the SM \cite{cao-octet}.
Also considering the different amplitude structure of Higgs single and pair production,
the cross section of Higgs pair production in the Manohar-Wise model
may deviate significantly from its predictions in the SM.

This paper is structured as follows.
In Sec.~II we briefly introduce the Manohar-Wise model. Then in Sec.~III
we present the numerical results and discussions of the Higgs pair production
in the Manohar-Wise model. Finally, the conclusions are presented in Sec.~IV.

\section{Model with color octet scalars ---the Manohar-Wise Model}
In the SM, the scalar sector contains only one Higgs scalar doublet,
which is responsible for the electroweak symmetry breaking.
Additional extensions of the scalar sector is restricted by
the principle of minimal flavor violation (MFV).
Just motivated by this principle, the Manohar-Wise model
extends the SM by adding one family color-octet scalars with
$SU(3)_C \times SU(2)_L \times U(1)_Y$ quantum numbers $(8,2)_{1/2}$ \cite{MW-Model},
\begin{eqnarray}
S^{A}=\left( \begin{array}{c}
        S^A_+ \\
        S^A_0
      \end{array} \right),
\end{eqnarray}
where $A=1,...,8$ denotes color index, $S^A_+$ and $S^A_0$ are the
electric charged and neutral color-octet scalar fields respectively, and
\begin{eqnarray}
S^A_0=\frac{1}{\sqrt{2}} (S^A_R+ i S^A_I)
\end{eqnarray}
with $S^A_{R, I}$ denote the neutral CP-even and CP-odd color-octet scalar fields.
In accordance with the MFV, the Yukawa couplings to the SM fermions
are parameterized as
\begin{eqnarray}
{\cal L} = -\eta_{U} Y_{ij}^U \bar{u}_R^i T^A S^A Q_L^j
- \eta_D Y_{ij}^D \bar{d}_R^i T^A (S^A)^\dagger Q_L^j + h.c.,
\end{eqnarray}
where $Y^{U,D}_{ij}$ are the SM Yukawa matrices, $i,j$ denote flavor
indices, and $\eta_{U,D}$ are flavor universal constants.

The most general renormalizable scalar potential is given by
\begin{eqnarray}
V&=& \frac{\lambda}{4} \big (H^{\dagger i} H_i - \frac{v^2}{2}\big
)^2 + 2m_S^2 \text{Tr} (S^{\dagger i}S_i) + \lambda_1 H^{\dagger
i}H_i \text{Tr} (S^{\dagger j}S_j) + \lambda_2 H^{\dagger i}H_j
\text{Tr} (S^{\dagger j}S_i)
\nonumber\\
&& + \big [ \lambda_3 H^{\dagger i}
H^{\dagger j} \text{Tr}(S_iS_j) + \lambda_4 H^{\dagger i}
\text{Tr}(S^{\dagger j}S_j S_i) +  \lambda_5 H^{\dagger i}
\text{Tr}(S^{\dagger j}S_i S_j) + h.c. \big ]
\nonumber\\
&& + \lambda_6 \text{Tr}
(S^{\dagger i}S_i S^{\dagger j} S_j) + \lambda_7
\text{Tr}(S^{\dagger i}S_j S^{\dagger j} S_i) + \lambda_8 \text{Tr}
(S^{\dagger i}S_i)\text{Tr}(S^{\dagger j}S_j)\8 \8 \nonumber\\
&& + \lambda_9 \text{Tr}
(S^{\dagger i} S_j) \text{Tr}(S^{\dagger j}S_i) + \lambda_{10}
\text{Tr} (S_i S_j) \text{Tr}(S^{\dagger i}S^{\dagger j})
+\lambda_{11} \text{Tr}(S_iS_j S^{\dagger j} S^{\dagger i}),
\end{eqnarray}
where $H$ is usual $(1,2)_{1/2}$ Higgs doublet, the traces are over color
indices with $S= S^A T^A$,
$i,j$ denote $SU(2)_L$ indices and all $\lambda_i$ ($i=1,..., 11$)
except $\lambda_4$ and $\lambda_5$ are real parameters.
Note that the convention $\lambda_3 >0$ is allowed by a appropriate
phase rotation of the $S$ fields.
After the electroweak symmetry breaking, the mass spectrum of the
scalars depend on the parameters in the scalar potential,
and at the tree-level are given by
\begin{eqnarray}
m_\pm^2 &=& m_S^2 + \lambda_1 \frac{v^2}{4}
          \equiv m_S^2 + \lambda_\pm \frac{v^2}{4},  \nonumber \\
m_R^2 &=& m_S^2 + (\lambda_1 + \lambda_2 + 2\lambda_3) \frac{v^2}{4}
        \equiv m_S^2 + 2\lambda_R \frac{v^2}{4}, \nonumber\\
m_I^2 &=& m_S^2 +(\lambda_1 + \lambda_2 - 2\lambda_3) \frac{v^2}{4}
        \equiv m_S^2 + 2\lambda_I \frac{v^2}{4}.   \label{mass}
\end{eqnarray}
The interactions of these scalars with the Higgs boson (labeled as $h$
denoting the SM Higgs boson) are as follows \cite{hgg-NLO},
\begin{eqnarray}
g_{hS^{A\ast}_i S^B_i} = \frac{v}{2} \lambda_i \delta^{AB},~~~~~~~
g_{hhS^{A\ast}_i S^B_i} = \frac{1}{2}\lambda_i \delta^{AB}  \label{hss}
\end{eqnarray}
with $i=\pm, R, I$, and we take $v=$ 246 GeV.

\section{Calculations and numerical results}

\begin{figure}[thbp]
\includegraphics[width=15cm]{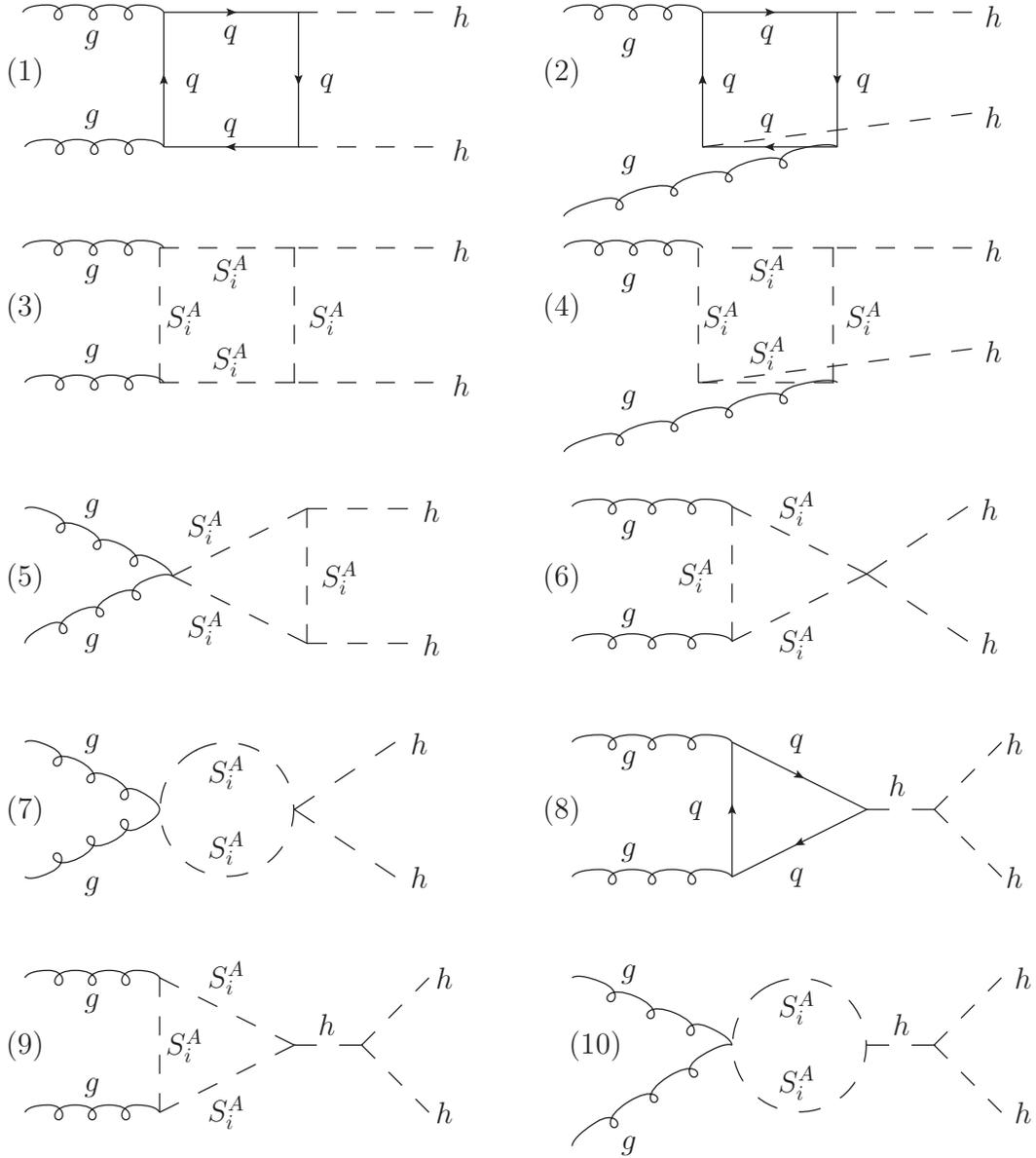}
\vspace{-0.5cm}
\caption{Feynman diagrams for the pair production of the Higgs boson
via gluon fusion in the Manohar-Wise model, with $S^A_i$ ($i=\pm, R, I$)
denoting the color-octet scalars in the model.
The diagrams with initial gluons or final Higgs bosons interchanged
are not shown here. Due to the large Yukawa couplings, we only consider the
contributions from the third generation quarks.}
\label{fig1}
\end{figure}
In the Manohar-Wise model the Higgs pair production at the LHC mainly proceeds
through the gluon fusion shown in Fig.\ref{fig1}.
Compared with the SM, the Manohar-Wise model predicts additional color octet scalars
including $S^A_i (i=\pm, R, I)$, which have couplings to the Higgs
boson $h$. Therefore, the pair production of $h$
in the Manohar-Wise model has additional contributions from the loops
of the color octet scalars $S^A_i (i=\pm, R, I)$ besides the
contributions from the loops mediated by the heavy quarks in the SM,
as shown in Fig.\ref{fig1}.
Since the additional contributions are at the same perturbation order
as those in the SM, the cross section of the Higgs pair production in
the Manohar-Wise model may significantly deviate from the prediction in the SM.

In the numerical calculations
we take $m_t=173$ GeV, $m_b=4.2$ GeV, $m_W=80.0$ GeV, $m_Z=91.0$ GeV,
and $\alpha=1/128$ \cite{PDG}, and fix the collision energy
of LHC and the mass of Higgs boson to be 14 TeV and 125.6 GeV respectively.
Then we use CT10 \cite{CT10} to
generate the parton distribution functions, with the factorization scale $\mu_F$
and the renormalization scale $\mu_R$ chosen to be $2 m_h$.
We check that the cross section of the Higgs pair production
in the SM is 18.7 fb, which is consistent with the result in \cite{sm-lo}.

In this work, following our previous work \cite{cao-octet}, we scan over
the parameter space of the Manohar-Wise model considering
following theoretical and experimental constraints:
(i) the constraints from the unitarity;
(ii) the constraints from electroweak precision data (EWPD);
(iii) the constraints from the LHC searches for exotic scalars through dijet-pair events.
Based on 4.6 fb$^{-1}$ data at 7-TeV LHC for dijet-pair events collected by the ATLAS collaboration,
the lower bound on the scalar mass has set to be 287 GeV at 95\% confidence level \cite{dijet}.
The lower bound from four-top channel is much higher, but it is based on some assumptions,
e.g., the bound is 500 GeV (630 GeV) for the neutral scalar decays into top pair with a
branching ratio of 50\% (100\%) \cite{four-top}. Since the latter constraint can be escaped from  by
adjusting $\eta_U$, we only require the color octet scalars to be heavier than 300 GeV.
Here we can comment that, in future running of the LHC the lower bound from dijet-pair events may be higher.
According to \cite{GoncalvesNetto:2012nt}, for a color-octet scalar of 350 GeV (500 GeV), its pair production rate
can reach 84.6 pb (11.4 pb) at 14-TeV LHC.

Under the above constraints, we perform fits in this model to the latest
Higgs data by using the ATLAS data and CMS data respectively.
The detail of the fits can be found in our previous works \cite{cao-octet, LH-hfit}.
From the fits we pick up the 1$\sigma$ (68\% confidence level or
$\chi^2_{min} \leq \chi^2 \leq \chi^2_{min} + 2.3$) and
2$\sigma$ (95\% confidence level or $\chi^2_{min} + 2.3  < \chi^2 \leq \chi^2_{min} + 6.18$) samples,
which correspond to $5.63 \leq \chi^2 \leq 7.93$ and $7.93 < \chi^2 \leq 11.81$ for the fit to the ATLAS data,
and $2.47 \leq \chi^2 \leq 4.77$ and $4.77 < \chi^2 \leq 8.65$ for the fit to the CMS data.
Then with these samples we calculate the cross section of Higgs pair production in the Manohar-Wise model
and define $R$ as the ratio normalized to its SM values,
\begin{eqnarray}
  R\equiv \sigma_{MW}(gg\to hh)/\sigma_{SM}(gg\to hh)
\end{eqnarray}

\begin{figure}
\includegraphics[width=15cm]{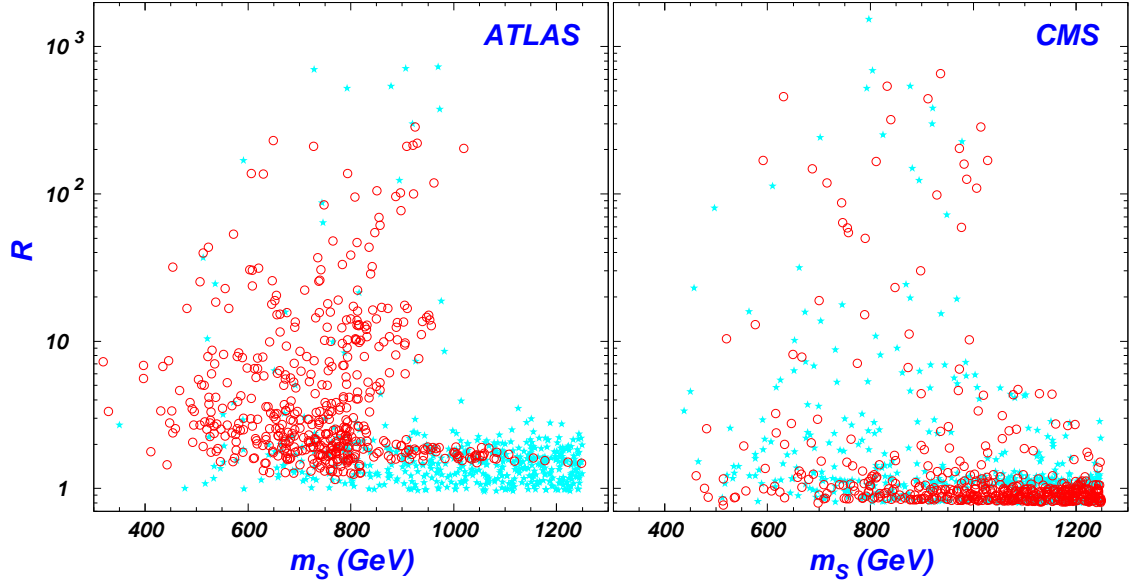}
\vspace{-0.5cm} \caption{The scatter plots of the surviving samples,
showing the normalized ratio $R$ as a function of $m_S$.
The red circles '$\circ$'  denote 1$\sigma$ surviving samples,
and the sky blue stars '$\star$' denote 2$\sigma$ samples.}
\label{fig2}
\end{figure}

\begin{figure}
\includegraphics[width=15cm]{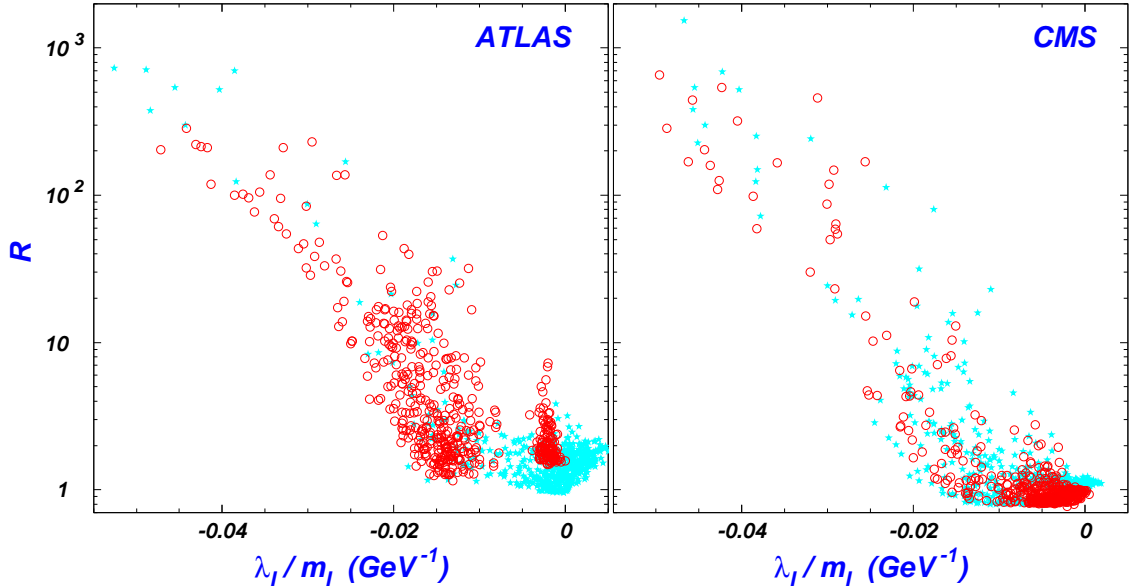}
\vspace{-0.5cm} \caption{Same as Fig.\ref{fig2}, but
showing the ratio of cross section in the Manohar-Wise model with that
in the SM (i. e. $R$) as a function of $\lambda_I/m_I$.}
\label{fig3}
\end{figure}

In Fig.\ref{fig2} we project the $1\sigma$ and $2\sigma$ samples on the plane of
the normalized ratio $R$ versus $m_S$.
The left panel displays the surviving samples in fitting to
the ATLAS Higgs data, and the right panel shows that to the CMS data.
In the figure, the red circles denote 1$\sigma$ surviving samples,
and the sky blue stars denote 2$\sigma$ samples.
From this figure we can clearly see that the cross section of the Higgs
pair production in the Manohar-Wise model can significantly deviate from the
SM prediction, and the normalized production rate $R$ can even be up to $10^3$.
The figure also shows that, for $m_S\gtrsim$ 1 TeV, the ratio $R$ is relatively small,
usually smaller than 10, which reflects the decoupling effect.

Now we give analytical explanations to the deviation of the
cross section in the Manohar-Wise model shown in Fig.\ref{fig2}.
The diagrams in Fig.\ref{fig1} can be divided into five parts:
(1)+(2), (3)+(4), (5), (6)+(7) and (8)+(9)+(10), and each part is
UV finite. We numerically check their relative size
and find that the contributions to the cross section from the
diagrams (3)+(4) and (5) are quite large. This is because the amplitude
of these diagrams can be written as
\begin{eqnarray}
\mathcal{M} &\sim & c_1\frac{g^2_{hS^{A\ast}_\pm S^A_\pm}}{m^2_\pm} +
                c_2\frac{g^2_{hS^{A\ast}_R S^A_R}}{m^2_R} +
                c_3\frac{g^2_{hS^{A\ast}_I S^A_I}}{m^2_I}  \label{eq1}
\end{eqnarray}
where $c_i$ (i=1, 2, 3) are ${\cal{O}}(1)$ coefficients.
Considering the couplings shown in Eq.(\ref{hss}), we rewrite the Eq.(\ref{eq1}) as
\begin{eqnarray}
\mathcal{M} &\sim & (c_1\frac{\lambda^2_\pm}{m^2_\pm} +
                c_2\frac{\lambda^2_R}{m^2_R} +
                c_3\frac{\lambda^2_I}{m^2_I})\frac{v^2}{4}     \label{eq2}
\end{eqnarray}
And the values of $\lambda_i$ ($i=\pm, R, I$) are usually large required by the
Higgs data \cite{cao-octet}.
While the amplitude from the other diagrams, such as (6)-(10) are not enhanced by $\lambda^2_i$
and usually proportional to $(C_{hgg}/SM)_i$ ,
whose summation can not diverge much from that of the SM,
since $|C_{hgg}/SM|\simeq 1$ according to current Higgs data (Fig. 2 in \cite{cao-octet}).
Besides, we also find that there are strong cancelation between the diagrams (3)+(4) and (5).

In our calculation, we find that the term involving
$\lambda^2_I/m^2_I$ are usually much larger than that of
$\lambda^2_{\pm}/m^2_{\pm}$ and $\lambda^2_R/m^2_R$ in Eq.(\ref{eq2}).
The reason can be understood as follows. Firstly, the surviving
samples prefer negative $\lambda_I$
and $|\lambda_I|$ is usually much larger compared with $\lambda_\pm$
and $\lambda_R$ (see Figure 1 in \cite{cao-octet}).
Secondly, Eq.(\ref{mass}) manifests that, for fixed $m_S$ and negative
$\lambda_i (i=\pm, R, I)$,
the larger $|\lambda_i|$, the smaller $m_i$.
Therefore, the contributions of the third term are dominant in Eq.(\ref{eq2}),
that is, the contributions from the loops mediated
by the scalar $S^A_I$ are much larger than that by the scalar $S^A_\pm$ and $S^A_R$.
As a proof, in Fig.\ref{fig3} we show the ratio $R$ versus $\lambda_I/m_I$.
The figure clearly shows that larger $|\lambda_I/m_I|$
usually predicts larger value of ratio $R$.
We checked that, for samples with $R\gtrsim 100$, the CP-odd octet scalars are 
not very light ($300\lesssim m_I \lesssim 600$ GeV), but the coupling $\lambda_I$ 
should be very large ($-25 \lesssim \lambda_I \lesssim -8$), which can also be understood 
from Figure 1 in \cite{cao-octet}. And these large-$R$ samples can also satisfy the 
perturbation theory, which suggests $|\lambda_i| \lesssim 8\pi$ ($i=\pm, R, I$)\cite{perturbation}.
Fig.\ref{fig3} also shows that for some special samples with $|\lambda_I/m_I|\sim 0$
in the left panel, the cross section in the Manohar-Wise model can also be enhanced
up to 10 times prediction in the SM.
For these samples, $|\lambda_R/m_R|$ is near 0.02 and the contributions
from Eq.(\ref{eq2}) can be still large,
comparable to that for the samples with $|\lambda_I/m_I|\sim 0.02$.
That can be understood from Figure 3 in \cite{cao-octet}.

\begin{figure}
\includegraphics[width=15cm]{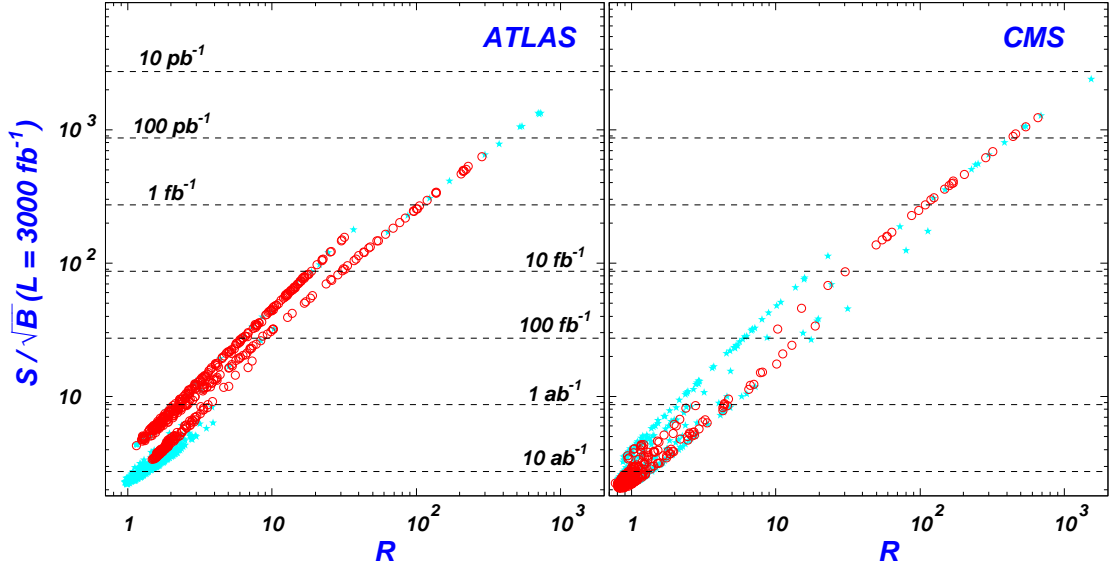}
\vspace{-0.5cm} \caption{Same as Fig.\ref{fig2}, but
showing the normalized ratio $R$ as a function of $S/\sqrt{B}$,
which is calculated at an integrated luminosity of 3000 fb$^{-1}$,
and also marking out the corresponding luminosity for $S/\sqrt{B}=5$.}
\label{fig4}
\end{figure}

Finally, we investigate the potential for discovery of Higgs pair production at the LHC14.
In Fig.\ref{fig4}, we project samples on the plane of
significance $S/\sqrt{B}$ versus the normalized ratio $R$.
In calculating $S/\sqrt{B}$, we utilize the Monte Carlo (MC) simulation result
of $gg\to hh \to b\bar{b}\gamma\gamma$ in the SM \cite{Yao:2013ika}.
We assume that in the Manohar-Wise model the $\sigma\times Br$
and acceptances of the background, the acceptances of the signal are the same as that in the SM,
while the $\sigma\times Br$ of the signal are calculated by ourselves,
which can be expressed as
\begin{eqnarray}
  (\sigma\cdot Br)_{MW} &=& \sigma_{SM} \times R \times BR(h\to b\bar{b}) \times Br(h\to \gamma\gamma) \nonumber \\
  &\simeq& (\sigma\cdot Br)_{SM} \times R \times (C_{h\gamma\gamma}/SM)^2,
\end{eqnarray}
thus $S/\sqrt{B}$ in the Manohar-Wise model should be proportional to
$R\times (C_{h\gamma\gamma}/SM)^2$.
So combined with Fig.2 and Fig.3 in \cite{cao-octet}, we can understand
that there are mainly three linear relation in each planes in Fig. \ref{fig4} in this paper.
Since $S/\sqrt{B}$ is also  proportional to $\sqrt{\emph{L}}$,
in this figure we also mark out the lines of $S/\sqrt{B}=5$ for other values of luminosity,
samples above which can be discovered with corresponding luminosity.
For example, with the integrated luminosity of 100 fb$^{-1}$ at the 14 TeV LHC,
when the cross section of Higgs pair production in the Manohar-Wise
model is enhanced by 10 times the prediction in the SM, i.e. $R=10$,
this process may be detected.
Owing to the highly enhanced Higgs pair production rate,
many samples in the Manohar-Wise model can be tested very soon after LHC running again.

\section{Summary and Conclusion}
Motivated by the principle of minimal flavor violation,
the Manohar-Wise model introduces one family of color-octet
scalars, which can have large couplings with the Higgs boson.
Since the properties of the SM-like Higgs boson around 125 GeV need to be precisely scrutinized,
in this work we studied the Higgs pair production considering the
effect of the color-octet scalars.
Following our previous work \cite{cao-octet}, we first scanned over the parameter space
of the Manohar-Wise model considering the theoretical and experimental constraints
and performed fits of the model to the latest Higgs data by using
the ATLAS and CMS data separately.
Then we calculated the Higgs pair production rate
and investigated the potential of its discovery at the LHC14.

Base on our calculation and analysis, we get following conclusions:
\begin{itemize}
  \item Under current constrains including Higgs data after Run I of the LHC,
  the cross section of Higgs pair production in the Manohar-Wise model
  can be enhanced up to even $10^3$ times prediction in the SM.
  \item Moreover, the sizable enhancement comes from the contributions
  of the CP-odd color-octet scalar $S^A_I$. For lighter scalar $S^A_I$ and
  larger values of $|\lambda_I|$, the cross section of Higgs pair production can be much larger.
  \item After running again of LHC at 14 TeV, most of the parameter spaces in the Manohar-Wise model can be test.
  For an integrated luminosity of 100 fb$^{-1}$ at the LHC14,
  when the normalized ratio $R=10$, the process of Higgs pair production can be detected.
\end{itemize}

\section*{Acknowledgement}
We thank Prof. Junjie Cao for helpful discussions.
This work was supported in part by the National Natural Science
Foundation of China (NNSFC) under grant No. 11247268, 11305050, and
by Specialized Research Fund for the Doctoral Program of Higher
Education with grant No. 20124104120001.



\begin{thebibliography}{32}
\bibitem{1207atlas}
  G.~Aad {\it et al.}  [ATLAS Collaboration],
  Phys.\ Lett.\ B {\bf 716} (2012) 1
  [arXiv:1207.7214 [hep-ex]].

\bibitem{1207cms}
  S.~Chatrchyan {\it et al.}  [CMS Collaboration],
  Phys.\ Lett.\ B {\bf 716} (2012) 30
  [arXiv:1207.7235 [hep-ex]].

\bibitem{13atlas}
The ATLAS Collaboration,
  ATLAS-CONF-2013-012;
  ATLAS-CONF-2013-034.

\bibitem{13cms}
The CMS Collaboration,
  CMS-PAS-HIG-13-001;
CMS-PAS-HIG-13-005.

\bibitem{125-our}
   J.~Cao {\it et al.},
  JHEP {\bf 1203}, 086 (2012) [arXiv:1202.5821 [hep-ph]];
JHEP {\bf 1210} (2012) 079
  [arXiv:1207.3698 [hep-ph]];
  Phys.\ Lett.\ B {\bf 710}, 665 (2012)
  [arXiv:1112.4391 [hep-ph]];
  Phys.\ Lett.\ B {\bf 703}, 462 (2011)
  [arXiv:1103.0631 [hep-ph]];
  JHEP {\bf 1206}, 145 (2012)
  [arXiv:1203.0694 [hep-ph]];
  Z.~Heng,
  Adv.\  High Energy Phys.\  {\bf 2012}, 312719 (2012)
  [arXiv:1210.3751 [hep-ph]].

\bibitem{125-other-MSSM}
  M.~Carena {\it et al.},
  JHEP {\bf 1203}, 014 (2012);
    A.~Arbey, M.~Battaglia, F.~Mahmoudi,
  Eur.\ Phys.\ J.\ C {\bf 72} (2012) 1906;  
S.~Heinemeyer, O.~Stal, G.~Weiglein,
  Phys.\ Lett.\ B {\bf 710} (2012) 201;  
  N. D. Christensen, T. Han, S. Su,
  Phys.\ Rev.\ D {\bf 85} (2012) 115018;  
  P. Lodone,
  Int.\ J.\ Mod.\ Phys.\ A {\bf 27} (2012) 1230010;  
  V.~Barger, M.~Ishida and W.~-Y.~Keung,
  Phys.\ Rev.\ D {\bf 87} (2013) 015003;  
  K. Hagiwara, J. S. Lee, J. Nakamura,
  JHEP {\bf 1210} (2012) 002;  
  F.~Boudjema and G.~D.~La Rochelle,
  Phys.\ Rev.\ D {\bf 86} (2012) 115007;
  P.~Bechtle {\it et al.},
  Eur.\ Phys.\ J.\ C {\bf 73} (2013) 2354;
  J.~Ke {\it et al.},  
  Phys.\ Lett.\ B {\bf 723} (2013) 113;
  K.~Cheung, C.~-T.~Lu and T.~-C.~Yuan,
  Phys.\ Rev.\ D {\bf 87} (2013) 075001;
    A.~Chakraborty {\it et al.},
  arXiv:1301.2745 [hep-ph];
  R.~S.~Hundi,
  Phys.\ Rev.\ D {\bf 87} (2013) 115005;
  T.~Han, T.~Li, S.~Su and L.~-T.~Wang,
  arXiv:1306.3229 [hep-ph];
  A.~Farzinnia, H.~-J.~He and J.~Ren,
  arXiv:1308.0295 [hep-ph].

\bibitem{natural-susy-125}
   L.~J.~Hall, D.~Pinner, J.~T.~Ruderman,
  JHEP {\bf 1204} (2012) 131;  
  A.~Arvanitaki, G.~Villadoro,
  JHEP {\bf 1202} (2012) 144;  
   J.~L.~Feng,
  arXiv:1302.6587 [hep-ph];
  K.~Kowalska and E.~M.~Sessolo,
  arXiv:1307.5790 [hep-ph];
  C.~Han {\it et al.},
  arXiv:1304.5724 [hep-ph];
  JHEP {\bf 1310}, 216 (2013)
  [arXiv:1308.5307 [hep-ph]].

\bibitem{NMSSM-125}
  U.~Ellwanger,
  JHEP {\bf 1203}, 044 (2012);
    U. Ellwanger, C. Hugonie,
  Adv.\ High Energy Phys.\  {\bf 2012} (2012) 625389;
    J.~F.~Gunion, Y.~Jiang, S.~Kraml,
  Phys.\ Lett.\ B {\bf 710} (2012) 454;
  JHEP {\bf 1210} (2012) 072;
  S.~F.~King {\it et al.},
  Nucl.\ Phys.\ B {\bf 860} (2012) 207;
  Nucl.\ Phys.\ B {\bf 870} (2013) 323;
   R. Benbrik {\it et al.},
   Eur.\ Phys.\ J.\ C {\bf 72} (2012) 2171;
  K.~Agashe, Y.~Cui and R.~Franceschini,
  JHEP {\bf 1302} (2013) 031;
  K.~Kowalska {\it et al.},
  Phys.\  Rev.\ D {\bf 87} (2013) 115010;
  L.~Aparicio {\it et al.},
  JHEP {\bf 1302} (2013) 084;
  T.~Gherghetta {\it et al.},
  JHEP {\bf 1302} (2013) 032;
  N.~D.~Christensen, T.~Han, Z.~Liu and S.~Su,
  JHEP {\bf 1308} (2013) 019;
  M.~Badziak {\it et al.},
  JHEP {\bf 1306} (2013) 043;
  T.~Cheng {\it et al.},
  arXiv:1304.3182 [hep-ph];
  Phys.\ Rev.\ D {\bf 88} (2013) 015031;
  S.~Moretti, S.~Munir and P.~Poulose,
  arXiv:1305.0166 [hep-ph].

\bibitem{cmssm-125}
H.~Baer {\it et al.},
  Phys.\ Rev.\ D {\bf 85}, 075010 (2012)
  [arXiv:1112.3017 [hep-ph]];
  JHEP {\bf 1205} (2012) 091;
  Phys.\ Rev.\ D {\bf 87} (2013) 3, 035017;
 J.~L.~Feng, K.~T.~Matchev and D.~Sanford,
  Phys.\ Rev.\ D {\bf 85} (2012) 075007;
 O.~Buchmueller {\it et al.},
  Eur.\ Phys.\ J.\ C {\bf 72} (2012) 2020;
  S.~Akula {\it et al.},
  Phys.\ Rev.\ D {\bf 85}, 075001 (2012);
  M.~Kadastik {\it et al.},
  JHEP {\bf 1205} (2012) 061;
  L.~Aparicio, D.~G.~Cerdeno, L.~E.~Ibanez,
  JHEP {\bf 1204} (2012) 126;
   J.~Ellis {\it et al.},
  Eur.\ Phys.\ J.\ C {\bf 72} (2012) 2005;
  Eur.\ Phys.\ J.\ C {\bf 73} (2013) 2403;
  Z.~Kang {\it et al.},  
  Phys.\ Rev.\ D {\bf 86} (2012) 095020;
  A. Fowlie {\it et al.},
  Phys.\ Rev.\ D {\bf 86} (2012) 075010;
  S. Akula, P. Nath, G. Peim,
  Phys.\ Lett.\ B {\bf 717} (2012) 188;
  O.~Buchmueller {\it et al.},
  Eur.\ Phys.\ J.\ C {\bf 72} (2012) 2243.

\bibitem{self-coupling}
  J.~Baglio {\it et al.},
  JHEP {\bf 1304}, 151 (2013)
  [arXiv:1212.5581 [hep-ph]];
D.~Y.~Shao, C.~S.~Li, H.~T.~Li and J.~Wang,
  arXiv:1301.1245 [hep-ph];
  M.~J.~Dolan, C.~Englert and M.~Spannowsky,
  Phys.\ Rev.\ D {\bf 87}, 055002 (2013).


\bibitem{hh-detect}
F.~Goertz, A.~Papaefstathiou, L.~L.~Yang and J.~Zurita,
  JHEP {\bf 1306}, 016 (2013)
  [arXiv:1301.3492 [hep-ph]];
  A.~Papaefstathiou, L.~L.~Yang and J.~Zurita,
  Phys.\ Rev.\ D {\bf 87}, 011301 (2013)
  [arXiv:1209.1489 [hep-ph]];
  J.~Baglio {\it et al.},
  JHEP {\bf 1304}, 151 (2013)
  [arXiv:1212.5581 [hep-ph]];
  M.~J.~Dolan, C.~Englert and M.~Spannowsky,
  JHEP {\bf 1210}, 112 (2012)
  [arXiv:1206.5001 [hep-ph]];
  N.~D.~Christensen, T.~Han and T.~Li,
  Phys.\ Rev.\ D {\bf 86}, 074003 (2012)
  arXiv:1206.5816 [hep-ph];
  R.~Contino {\it et al.},
  JHEP {\bf 1208}, 154 (2012)
  arXiv:1205.5444 [hep-ph];
  U.~Baur, T.~Plehn and D.~L.~Rainwater,
  Phys.\ Rev.\ D {\bf 69}, 053004 (2004)
  [hep-ph/0310056];
  N.~Haba, K.~Kaneta, Y.~Mimura and E.~Tsedenbaljir,
  arXiv:1311.0067 [hep-ph];
  X.~Li and M.~B.~Voloshin,
  arXiv:1311.5156 [hep-ph].

\bibitem{Barger:2013jfa}
  V.~Barger, L.~L.~Everett, C.~B.~Jackson and G.~Shaughnessy,
  arXiv:1311.2931 [hep-ph].

\bibitem{hh-other}
  J.~-J.~Liu {\it et al.},
  Phys.\ Rev.\ D {\bf 70}, 015001 (2004);
  L.~Wang and X.~-F.~Han,
  Phys.\ Lett.\ B {\bf 696}, 79 (2011);
  X.~-F.~Han, L.~Wang and J.~M.~Yang,
  Nucl.\ Phys.\ B {\bf 825}, 222 (2010);
  L.~Wang {\it et al.},
  Phys.\ Rev.\ D {\bf 76}, 017702 (2007);
    H.~Sun {\it et al.},
  Eur.\ Phys.\ J.\  {\bf 72}, 2011 (2012);
  arXiv:1211.6201.

\bibitem{hh-susy}
J.~Cao, Z.~Heng, L.~Shang, P.~Wan and J.~M.~Yang,
  JHEP {\bf 1304}, 134 (2013);
Z.~Heng, L.~Shang and P.~Wan,
  JHEP {\bf 1310}, 047 (2013);
U.~Ellwanger,
  JHEP {\bf 1308}, 077 (2013)
  [arXiv:1306.5541 [hep-ph]];
C.~Han, X.~Ji, L.~Wu, P.~Wu and J.~M.~Yang,
  arXiv:1307.3790 [hep-ph];
J.~Cao, Y.~He, P.~Wu, M.~Zhang and J.~Zhu,
  arXiv:1311.6661 [hep-ph].


\bibitem{color-scalar}
G.~D.~Kribs and A.~Martin,
  Phys.\ Rev.\ D {\bf 86}, 095023 (2012)
  [arXiv:1207.4496 [hep-ph]];
  S.~Dawson, E.~Furlan and I.~Lewis,
  Phys.\ Rev.\ D {\bf 87}, 014007 (2013)
  [arXiv:1210.6663 [hep-ph]];
  T.~Enkhbat,
  arXiv:1311.4445 [hep-ph];
I.~Dorsner, S.~Fajfer, A.~Greljo and J.~F.~Kamenik,
  JHEP {\bf 1211}, 130 (2012)
  [arXiv:1208.1266 [hep-ph]].

\bibitem{MW-Model}
  A.~V.~Manohar and M.~B.~Wise,
  Phys.\ Rev.\ D {\bf 74} (2006) 035009
  [hep-ph/0606172].

\bibitem{cao-octet}
  J.~Cao, P.~Wan, J.~M.~Yang and J.~Zhu,
  JHEP {\bf 1308}, 009 (2013)
  [arXiv:1303.2426 [hep-ph]].

\bibitem{hgg-NLO}
  R.~Bonciani, G.~Degrassi and A.~Vicini,
  JHEP {\bf 0711} (2007) 095
  [arXiv:0709.4227 [hep-ph]].

\bibitem{PDG}
  J.~Beringer {\it et al.}  [Particle Data Group Collaboration],
  Phys.\ Rev.\ D {\bf 86}, 010001 (2012).

\bibitem{CT10}
  H.~-L.~Lai {\it et al.},
  Phys.\ Rev.\ D {\bf 82}, 074024 (2010)  arXiv:1007.2241 [hep-ph].

\bibitem{sm-lo}
  A.~Djouadi, W.~Kilian, M.~Muhlleitner and P.~M.~Zerwas,
  Eur.\ Phys.\ J.\ C {\bf 10}, 45 (1999).

\bibitem{dijet}
G.~Aad {\it et al.}  [ATLAS Collaboration],
  Eur.\ Phys.\ J.\ C {\bf 71}, 1828 (2011)
  [arXiv:1110.2693 [hep-ex]];
  Eur.\ Phys.\ J.\ C {\bf 73}, 2263 (2013)
  [arXiv:1210.4826 [hep-ex]].

\bibitem{four-top}
ATLAS collaboration,
ATLAS-CONF-2012-130 (2012);
ATL-PHYS-PROC-2013-016.

\bibitem{GoncalvesNetto:2012nt}
  D.~Goncalves-Netto, D.~Lopez-Val, K.~Mawatari, T.~Plehn and I.~Wigmore,
  Phys.\ Rev.\ D {\bf 85}, 114024 (2012)
  [arXiv:1203.6358 [hep-ph]].

\bibitem{LH-hfit}
  X.~-F.~Han, L.~Wang, J.~M.~Yang and J.~Zhu,
  Phys.\ Rev.\ D {\bf 87} (2013) 5,  055004
  [arXiv:1301.0090];
  L.~Wang, J.~M.~Yang and J.~Zhu,
  Phys.\ Rev.\ D {\bf 88} (2013) 075018
  [arXiv:1307.7780 [hep-ph]].

\bibitem{perturbation}
  S.~Kanemura, T.~Kasai and Y.~Okada,
  Phys.\ Lett.\ B {\bf 471}, 182 (1999)
  [hep-ph/9903289];
  E.~Accomando, {\it et al.},
  hep-ph/0608079.

\bibitem{Yao:2013ika}
  W.~Yao,
  arXiv:1308.6302 [hep-ph].

\end{thebibliography}
\end{document}